\documentclass[a4paper,11pt]{article}

\usepackage[utf8x]{inputenc}
\usepackage{amsmath}
\usepackage{amssymb}
\usepackage{graphicx}
\usepackage{jheppub}
\usepackage{bbm}
\usepackage{braket}

\newcommand{\df}{\mathrm{d}}

\title{Glauber Gluons and Multiple Parton Interactions}
\author{Jonathan R. Gaunt}

\affiliation{Theory Group, Deutsches Elektronen-Synchrotron (DESY), D-22607 Hamburg, Germany}

\emailAdd{jonathan.gaunt@desy.de}

\abstract{ 
We show that for hadronic transverse energy $E_T$ in hadron-hadron collisions, 
the classic Collins-Soper-Sterman (CSS) argument for the cancellation of Glauber gluons breaks down at
the level of two Glauber gluons exchanged between the spectators. Through an argument that 
relates the diagrams with these Glauber gluons to events containing additional soft scatterings,
we suggest that this failure of the CSS cancellation actually corresponds to a failure of the
`standard' factorisation formula with hard, soft and collinear functions to describe $E_T$
at leading power. This is because the observable receives a leading power contribution from multiple 
parton interaction (or spectator-spectator Glauber) processes. We also suggest that the same argument 
can be used to show that a whole class of observables, which we refer to as MPI sensitive observables, 
do not obey the standard factorisation at leading power. MPI sensitive observables are observables whose 
distributions in hadron-hadron collisions are disrupted strongly by the presence of multiple parton interactions 
(MPI) in the event. Examples of further MPI sensitive observables include the beam thrust $B^+_{a,b}$ and 
transverse thrust.
}

\begin{document}

{\flushright DESY 14-067\\{May 8, 2014}\\[-9ex]}

\maketitle

\section{Introduction}

Since the LHC is a proton-proton collider, collisions at the LHC are necessarily
a QCD-rich environment. A key component in making predictions at the LHC are 
factorisation formulae, which separate out the short-distance interaction we are
interested in from the long-distance QCD physics (pieces of which we may not be
able to calculate perturbatively, but are universal -- e.g. PDFs, fragmentation
functions, etc.). However, factorisation has only been rigorously proven in hadron-hadron 
collisions for the inclusive cross section $p + p \to V + X$, where $V$ is some hard final 
state and $X$ can be anything, as well as for the $p_T$ distribution of $V$, when $V$ is 
colourless \cite{Bodwin:1984hc, Collins:1985ue, Collins:1988ig,Collins:1350496}. These 
factorisation formulae are true at least to the leading power in $\Lambda_{QCD}/Q$ (where 
$Q$ is the hard scale associated with the production of $V$).

In the standard type of factorisation formula one has hard, collinear and (central) soft functions
describing high virtuality particles, particles approximately collinear to some beam or
jet direction, and particles with all components of momentum small and of the same order
respectively. Examples of collinear functions include PDFs, fragmentation functions, 
transverse momentum dependent PDFs (TMD PDFs) \cite{Collins:1984kg, Catani:2000vq, Collins:1350496, Becher:2010tm, Catani:2010pd, GarciaEchevarria:2011rb, Chiu:2012ir, Echevarria:2012js, Becher:2012yn}, 
and beam functions \cite{Stewart:2009yx, Stewart:2010qs}.
Another momentum region that can potentially contribute at the leading power
is the Glauber region. Let us decompose a general momentum $A$ according to
$A = A^+ p + A^- n + \mathbf{A}_T$, where $p.n = 1, n^2 = p^2 = 0$ and $\mathbf{A}_T$ is a 
vector perpendicular to $n$ and $p$. Then a Glauber momentum $r$ satisfies $|r^+r^-| \ll \mathbf{r}_T^2 \ll Q^2$. In order to obtain a `standard' factorisation
formula one has to show that the effects of the Glauber region cancel.  In the proofs of factorisation for the inclusive
total cross section and $p_T$ of $V$ \cite{Bodwin:1984hc,Collins:1985ue, Collins:1988ig,Collins:1350496},
the effect of the Glauber region is shown to cancel using an argument that relies on us
being able to sum inclusively over the undetected particles $X$ (equivalently it relies on us being
able to sum over the cuts of graphs containing Glauber gluons -- see below). We shall refer to 
this argument as the Collin-Soper-Sterman (CSS) argument \cite{Collins:1988ig} (see also \cite{Aybat:2008ct}).
In the following we shall also say that an observable factorises if we can write down a standard
`hard $\otimes$ collinear $\otimes$ central soft' factorisation formula for it, and for brevity write
`factorisation formula' rather than `standard factorisation formula'\footnote{We leave open the possibility for the observables we study in this paper, which do not satisfy the standard
factorisation, that a more general type of leading power factorisation incorporating Glauber modes may be possible.}.

A number of leading power factorisation formulae in hadron-hadron collisions have been written down 
for other observables, and some partially proven (especially in the context of soft-collinear effective field 
theory (SCET) \cite{Bauer:2000ew,Bauer:2000yr,Bauer:2001ct,Bauer:2001yt,Bauer:2002nz,Beneke:2002ph}.
Typically these proofs are limited in the sense that they only apply in the absence of Glauber modes
-- but see \cite{Idilbi:2008vm, D'Eramo:2010ak, Ovanesyan:2011xy, Bauer:2010cc, StewartTalk1, StewartTalk2, RothsteinTalk, Fleming:2014rea} for progress
in incorporating Glauber modes into SCET.

In this paper we investigate in detail the effect of Glauber gluons on, and factorisation 
properties of, two observables  -- hadronic transverse energy $E_T$, and beam thrust $B^+_{a,b}$, when 
the hard process is the production of a colourless particle $V$ ($V = H,W,Z...$) with associated scale $Q$. To 
obtain the $E_T$ in the event $pp \to V + X$ we take every particle $i$ in $X$, compute 
$\sqrt{m_i^2 + \mathbf{p}_{Ti}^2}$, and then sum over $i$. To define beam thrust we divide 
the event into two hemispheres, one of which contains $p$ and the other of which contains $n$ 
(with $p$ oriented along one proton direction and $n$ oriented along the other). In the $p$ 
hemisphere we take each non-$V$ particle $i$ with momentum $p_i$, compute $\sqrt{2}p.p_i$, and 
then sum over $i$ to obtain $B_a^+$. To obtain $B_b^+$ we look in the $n$ hemisphere and sum 
over $\sqrt{2}n.p_i$. A standard factorisation formula for the $E_T$ distribution has been written
down in \cite{Papaefstathiou:2010bw, Tackmann:2012bt, Grazzini:2014uha} (although in \cite{Papaefstathiou:2010bw, Grazzini:2014uha}
it is referred to as a resummation formula), and a standard factorisation
formula for $B_{a,b}^+$ has been written down in \cite{Stewart:2009yx}. In the latter case, additional 
arguments were given to rule out possible  Glauber effects, relying in part on the original CSS arguments 
\cite{Collins:1988ig, Aybat:2008ct}. Our analysis below indicates that the given arguments are not sufficient to rule out Glauber 
contributions from spectator-spectator interactions, which as we will see are related to multiparton 
interactions (MPI). When one makes plots of these observables using Monte Carlo generators, 
one observes a significant impact on their shapes arising from MPI
in the underlying event (UE) \cite{Papaefstathiou:2010bw, Grazzini:2014uha, AlcarazMaestre:2012vp}, 
and the resulting shapes look very much broader than the predictions from the factorisation formula. 
One might take this as a suggestion that the standard factorisation formulae may be inadequate for these 
observables, and this is one reason why we choose to re-visit their factorisation properties.

To begin with, in section \ref{sec:ET}, we will just consider the $E_T$
case in detail. We will see if the CSS argument (or some slight modification thereof), which functioned 
successfully for the inclusive cross section and $p_T$ cases, can also be applied to the observable $E_T$,
when this variable is parametrically smaller than the hard scale $Q$. What we shall see is that a slight 
modification of the argument works at the level of one Glauber gluon exchange, but fails when we have two 
Glauber gluon exchanges. Our discussion very closely follows that of section 14.3 of \cite{Collins:1350496}, 
and involves using the Libby-Sterman analysis \cite{Libby:1978bx, Sterman:1978bi}. This approach involves 
identifying pinch surfaces of Feynman graphs and then using power counting to identify the pinches that 
contribute to leading power. As part of our discussion we will review the cancellation of Glauber gluon effects 
according to the CSS argument for the observables $p_T$ of $V$ and the total cross section. 

We will also highlight the connection between this type of two Glauber gluon exchange graph and the
soft MPI modelled in Monte Carlo generators. The connection between Glauber gluons and soft MPI has been
made before \cite{Collins:1350496, Forshaw:2012bi} but here we devote it particular attention.
Using this connection we will suggest that the failure of the CSS argument strongly indicates a
breakdown in the standard factorisation.

Due to the connection between Glauber interactions and MPI, there are problems with applying the 
standard factorisation formula for a wider class of observables. We refer to these variables as 
MPI sensitive observables and choose the symbol $O_{S}$ to refer to a generic MPI sensitive observable.
In section \ref{sec:OtherObs} we discuss some of these observables, including beam thrust $B^+_{a,b}$
and transverse thrust. We also discuss how jet-based observables are less MPI sensitive than global 
observables such as $E_T$ and $B^+_{a,b}$.

Our main discussion is for the case where $O_S$ is parametrically smaller than the hard scale $Q$
(where here we assume always that $O_S$ has mass dimension $1$ -- if this is not the case then the
statements below have to be altered in a straightforward manner). We shall however devote some discussion 
to the case $O_S \sim Q$ in section \ref{sec:ObsSimHard}. 
Here one might expect the factorisation breaking effects not to be so significant, as the soft Glauber 
scattering/additional low-scale interaction affects the observable comparatively less. However, since 
multiple additional scatterings are possible (and tend to occur) and have a cumulative effect on 
$O_S$, there is the possibility of the Glauber gluons/MPI having a significant effect on these observables even 
when they are of order $Q$. As an example we shall discuss the case of $E_T$, where detailed Monte 
Carlo studies have been performed \cite{Papaefstathiou:2010bw, Grazzini:2014uha} and show a strong
disruption of the $E_T$ spectrum up to values of order $Q$ and beyond, when $Q$ was of order $100$ GeV.
This discussion will be of a somewhat less formal nature than that in sections \ref{sec:ET} and 
\ref{sec:OtherObs}.

\section{Violation of CSS Factorisation for MPI sensitive observables}

\subsection{Hadronic Transverse Energy} \label{sec:ET}

In this section we will demonstrate in detail that the CSS method of cancelling the Glauber region by 
summing over cuts for the Glauber subgraph (or some straightforward extension of the CSS method)
fails for the `MPI sensitive' observable $E_T$. At the same time we will review the success of 
the method for $p_T$ and the total cross section.

For our calculation we will use the model discussed in section 14.3 of \cite{Collins:1350496}.
In particular, we will take the proton to be composed of only two constituents
(say a quark and an antiquark). In our discussion we will assume very little about
the coupling of the constituents to the proton. In many past calculations the coupling
of the partons to the proton is taken to be `soft' in the sense that it suppresses large
transverse momenta of the outgoing constituents more strongly than the QCD vertex 
(such that there is no UV divergence at this coupling for large $k_T$) -- see for example
section 2.7 of \cite{Curci:1980uw}, or any of \cite{Paver:1982yp,Basu:1984ba, Gaunt:2012dd}.
If one takes this model for the coupling, then in the parlance of \cite{Gaunt:2012dd} the
quark and antiquark are a `nonperturbatively generated' pair, and this scenario is most 
appropriate when $E_T$ is of order $\Lambda_{QCD}$. Alternatively we could replace the
proton by a gluon from the proton, and take the
quark-antiquark-`proton' coupling to be the QCD coupling. Then the quark-antiquark pair
is `perturbatively generated', and this scenario is more appropriate when $E_T$ is perturbative.
We will denote the coupling of the constituents to the `proton' by a grey blob.

\begin{figure} 
\centering
\includegraphics[scale=0.6]{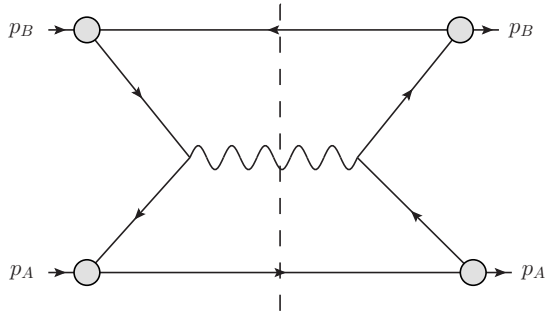}
\caption{Parton-model graph for the production of $V$ from the two hadrons $P_A$
and $P_B$. The only physical cut of this diagram is denoted on the graph using 
a dashed line.}
\label{fig:partonmodel}
\end{figure}

 We will
take the hard process to be $q\bar{q} \to V$ where the hard scale associated with
this is $Q$, and to avoid the complexities associated with final-state colour, we 
take $V$ to be colourless. Then the lowest order 
`parton model' process for $P_A + P_B \to V + X$ is shown in figure \ref{fig:partonmodel}.
For simplicity we take all partons (gluons + quarks) in the calculation to be 
massless. We take the spatial momentum of $P_A$ to point along that of $p$, and the
spatial momentum of $P_B$ to point along that of $n$. Protons $A$ and $B$ are taken to
have a small mass such that $P_A^-$ and $P_B^+$ are nonzero but small compared to 
$P_A^+$ and $P_B^-$ (this is not essential to our argument).

\begin{figure} 
\centering
\includegraphics[scale=0.6]{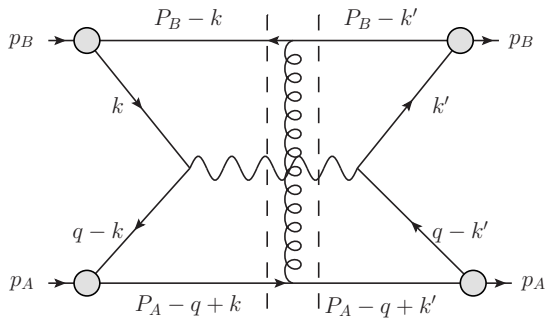}
\caption{Graph with one Glauber exchange between the spectators. We use the same
momentum assignments as figure 14.4 in \cite{Collins:1350496}. The two cuts that
allow the gluon to remain in the Glauber region are denoted on the graph as dashed
lines.}
\label{fig:oneglauber}
\end{figure}

We start with the most simple type of diagram that contains a Glauber gluon, 
which is the diagram in figure \ref{fig:oneglauber}. In this diagram we have a Glauber gluon
attached to two `final-state' spectators (where we use quotation marks because, 
as has been pointed out by Collins \cite{Collins:1350496} and as we shall review later, this is not an
interaction in the final state). Note that in this paper we will not consider
Glauber gluon attachments between the two initial state lines (i.e. the lines 
leading into the hard interaction) and between the initial state lines and the
spectator lines. It is well known that these type of interactions do not have 
a pinch in the Glauber region \cite{Collins:1350496, Collins:1997sr, Collins:2004nx}, so these types of interactions are handled
satisfactorily using the usual collinear and soft functions and there is no need to
study them further.

\begin{figure} 
\centering
\includegraphics[scale=0.6]{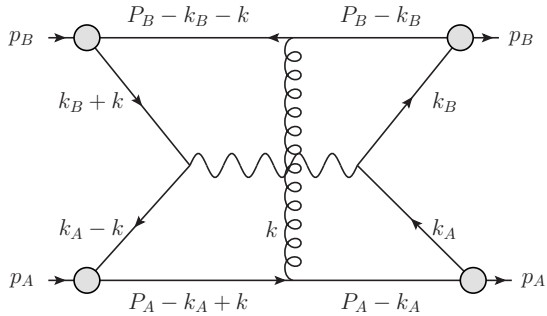}
\caption{Parameterisation of loop momentum in one-Glauber diagram that is particularly convenient
to show the Glauber pinch.}
\label{fig:oneglauberpinch}
\end{figure}

It is well known and easy to show that this diagram has a leading pinch singularity 
with the gluon exchanged between the spectators being trapped in the Glauber region
(leading means that this diagram contributes at the same power in $\Lambda_{QCD}/Q$
as the `parton model' diagram of figure \ref{fig:partonmodel}) \cite{Collins:1350496}. 
Let us take the parameterisation of the loop momenta to be as in figure \ref{fig:oneglauberpinch}, with $k$
as the Glauber momentum. We take $k_A$ and $k_B$ to be collinear to $P_A$ and $P_B$ 
-- i.e. their momenta are of order $(1,\lambda^2,\lambda)Q$ and $(\lambda^2,1,\lambda)Q$
respectively (where $\lambda$ is a `normal coordinate' that describes the degree of 
collinearity of the lines). Let us take the transverse component of $k$ to be of 
order $\lambda_SQ$. Which constraints do we have on $k^+$ and $k^-$? The 
structure of the integral over these components is as follows:
\begin{align}
\int &\dfrac{\df k^+ \df k^-}{(2\pi)^2} \dfrac{\text{numerator}}{2k^+k^- - \mathbf{k}_T^2+i0}
\\ \nonumber
& \times \dfrac{1}{[-2k^+(P_B^- - k_B^-) + ... + i0][2k^+k_B^-+...+i0]}
\\ \nonumber
& \times \dfrac{1}{[-2k^-k_A^+ + ... + i0][2k^-(P_A^+-k_A^+)+...+i0]}
\end{align}
where the terms indicated by ``...'' are all independent of $k^+$ and $k^-$ and of order
$Q(\lambda^2,\lambda\lambda_S, \lambda_S^2)$. We see that if $\lambda_S$ is of order $\lambda$
then $k^+$ and $k^-$ are trapped at small values of order $\lambda_S^2Q$ -- this is the 
Glauber pinch. The region around this pinch is leading (for $\lambda_S \sim \lambda$) -- 
compared to the leading parton model graph of figure \ref{fig:partonmodel} we have an 
enhancement of $1/\lambda^6$ from the extra 3 propagator denominators (there are no extra
factors of $\lambda$ from the numerator), plus a suppression from the (Glauber) phase space 
of $k$ which is of order $\lambda^6$ ($\lambda^2$ from each of $\df k^+$ and $\df k^-$, and $\lambda^2$
from $\df^2 \mathbf{k}_T$. This gives $\lambda^0$ overall -- i.e. leading. For more detail see \cite{Collins:1350496}.

\begin{figure} 
\centering
\includegraphics[scale=0.5]{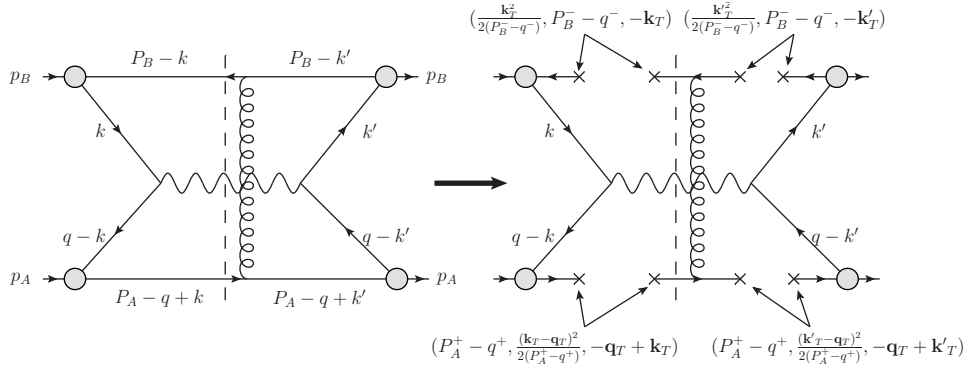}
\caption{Leading power decomposition of one of the cut graphs in figure \ref{fig:oneglauber} --
the decomposition for the other cut is the same, except with the cut on the right hand side of the
Glauber gluon. The lines with crosses on are now on-shell lines - we write the $(+,-,T)$ components
of these momenta near the lines.}
\label{fig:oneglaubercut}
\end{figure}

There are two possible cuts of this diagram which leave the gluon in the Glauber 
region, one to the left of the Glauber gluon, and one to the right. These are depicted using dashed lines
in figure \ref{fig:oneglauber}. Of course we could also have a cut that runs through the `Glauber' gluon 
itself, but then the gluon would be forced into the central soft region by the 
on-shellness condition, and would therefore no longer be in the region we want to 
consider here. It was shown in \cite{Collins:1350496} in the leading power approximation that 
for each cut of figure \ref{fig:oneglauber}, we can replace the momentum in the two spectator lines 
that are not already on shell by an on-shell momentum. This replacement is given 
pictorially for one of the cuts in figure \ref{fig:oneglaubercut}. 

Let us briefly review this argument. In the region we are considering (close to the pinch, in which
the lines connected to $P_A$ are approximately collinear to $P_A$, whilst the lines connected to $P_B$
are approximately collinear to $P_B$) the components $k^+$ and $k'^+$ are small, and we can neglect them 
in the bottom half of the graph compared to the large momentum components $q^+$ and $P_A^+-q^+$. Similarly,
in the top half of the graph we can neglect $q^- - k^-$ and $q^- - k'^-$ compared to the large components
$q^-$ and $P_B^- - q^-$. Finally, we can neglect the Glauber-trapped components $k^\pm$ and $k'^\pm$ compared to the larger components 
$\mathbf{k}_T$ and $\mathbf{k'}_T$.

Now let us consider the $k^+$ integration. After the approximations, the only dependence on $k^+$ in this
integration is in the lines $k$ and $P_B-k$, and we can perform the integral by closing the $k^+$ contour
on the pole of $(P_B -k)^2$:

\begin{align}
\int &\dfrac{\df k^+}{2\pi} \dfrac{i}{(2q^-k^+-\mathbf{k}_T^2+i0)}\dfrac{i}{(2(-k^++P_B^+)(P_B^--q^-)-\mathbf{k}_T^2+i0)}
\\  \nonumber
& = \dfrac{i}{2(P_B^--q^-)}\dfrac{i}{2q^-k^+_{\text{on-shell}}-\mathbf{k}_T^2+i0}
\\  \nonumber
& = \int \dfrac{\df k^+}{2\pi} \dfrac{i}{(2q^-k^+-\mathbf{k}_T^2+i0)}2\pi\delta(2(-k^++P_B^+)(P_B^--q^-)-\mathbf{k}_T^2+i0)
\end{align}
where
\begin{align}
k^+_{\text{on-shell}} = P_B^+ - \dfrac{\mathbf{k}_T^2}{2(P_B^--q^-)}
\end{align}

In the above we have ignored the numerator factor which is not relevant for the present discussion. The effect of the
integral over $k^+$ is to set $P_B-k$ on shell.
The same argument can be repeated for the integrations over $k'^+, k^-$ and $k'^-$ to set
the lines $P_B-k'$, $P_B-q+k$ and $P_B-q+k'$ on shell.

We can therefore separate the sum of two cut graphs into two 
factors, one of which is just the `parton model' graph with no Glauber gluon (but 
with slightly different momenta to the left and right of the cut, for general 
$\mathbf{k}_T, \mathbf{k'}_T$), and the second of which is the sum over the 
two external cuts for the $q\bar{q} \to q\bar{q}$ $t$-channel gluon exchange 
diagram. These two factors are integrated over $\mathbf{k}_T, \mathbf{k'}_T$,
together with any measurement delta functions.

Let's first say that we make no measurement on the final state (i.e. we are 
interested only in the fully inclusive cross section for the production of $V$). 
Then there are no measurement delta functions to worry about. Now, the gluon 
exchange graph with the external cut to the right of the graph is just the graph itself
$L(\mathbf{k}_T \to \mathbf{k}'_T)$. On the other hand, the graph with the external cut 
to the left is the complex conjugate graph for the time reversed process, $L^*(\mathbf{k}'_T 
\to \mathbf{k}_T)$. Graphs $L$ and the corresponding amplitudes $\mathcal{M}$ are related by
$\mathcal{M} = -iL$, so at the amplitude level we have $i\mathcal{M}(\mathbf{k}_T \to \mathbf{k}'_T)
-i\mathcal{M}^*(\mathbf{k}'_T \to \mathbf{k}_T)$. This is equal to minus the sum over internal cuts of 
the graph according to the Cutkosky rules \cite{Cutkosky:1960sp, Veltman:1994Ve}, which are the
generalised form of the optical theorem described in Appendix \ref{sec:GenOp}. However there are no
internal cuts of the graph, so the sum of external cuts gives zero and the effect of the Glauber gluons cancels at this order 
for the total cross section.

Now let's consider what happens when we add a measurement. What we would ideally
like to happen in order to cancel the Glauber diagrams is, that for fixed momenta 
$\mathbf{k}_T, \mathbf{k'}_T$, the two cuts have the same value of the measurement.
Then the two cuts of the gluon exchange diagram with the same momenta contribute 
to the same measurement, and we can add them together to give zero as for the total
cross section. This certainly works if the measurement is the $p_T$ of $V$ (or 
equivalently the $p_T$ of all other final state particles) -- the core reason behind
this is simply momentum conservation, which means that for given $\mathbf{k}_T, \mathbf{k'}_T$
the total $p_T$ of all final state particles not including $V$ is equal both `before' 
and `after' the gluon exchange.

On the other hand, this does not work for our `MPI sensitive' variable, the total $E_T$.
There the $E_T$ is $|\mathbf{k}_T| + |\mathbf{q}_T - \mathbf{k}_T|$ for the cut to the
left of the Glauber gluon, and $|\mathbf{k'}_T| + |\mathbf{q}_T - \mathbf{k'}_T|$ for 
the cut to the right. However we can still attempt to arrange a cancellation of the 
Glauber region at this simplest order as follows. We write the sum over two cuts as
follows:
\begin{align} \label{eq:1Gcancel}
& \int \df^{d-2}\mathbf{k}_T \, \df^{d-2}\mathbf{k'}_T  \, f_P(\mathbf{k}_T) f^*_P(\mathbf{k'}_T) L(\mathbf{k}_T \to \mathbf{k'}_T) \delta(E_T = |\mathbf{k'}_T|+|\mathbf{k'}_T - \mathbf{q}_T |) +
\\   \nonumber
& \int \df^{d-2}\mathbf{k}_T \, \df^{d-2}\mathbf{k'}_T \, f_P(\mathbf{k}_T) f^*_P(\mathbf{k'}_T) L^*(\mathbf{k'}_T \to \mathbf{k}_T) \delta(E_T = |\mathbf{k}_T|+|\mathbf{k}_T - \mathbf{q}_T |)
\end{align}
where $f_P$ is half the parton model graph (the piece to the left of the cut in figure \ref{fig:partonmodel}),
and $L$ is the one-Glauber exchange graph. We have suppressed a dependence on $P_A$, $P_B$, and $q$ in these 
factors to avoid too many function arguments.

In the second term we re-label $\mathbf{k}_T \leftrightarrow \mathbf{k'}_T$, and then combine
the terms, obtaining:
\begin{align} \label{eq:1Gcancel2}
 \int &\df^{d-2}\mathbf{k}_T \, \df^{d-2}\mathbf{k'}_T  \, \delta(E_T = |\mathbf{k'}_T|+|\mathbf{k'}_T - \mathbf{q}_T |) \times \\ \nonumber
&\left[ f_P(\mathbf{k}_T) f^*_P(\mathbf{k'}_T) L(\mathbf{k}_T \to \mathbf{k'}_T) + f_P(\mathbf{k'}_T) f^*_P(\mathbf{k}_T) L^*(\mathbf{k}_T \to \mathbf{k'}_T)\right]
\end{align}

There are no physical cuts of $f_P$ -- therefore $f_P^* = -f_P$ using the Cutkosky rules. 
Also to leading power the gluon exchange graph $L$ only depends on the transverse momenta through 
the gluon propagator denominator, which is $\sim (\mathbf{k}_T - \mathbf{k'}_T)^2$ at leading power.
$L$ is therefore symmetric in $\mathbf{k}_T$ and $\mathbf{k'}_T$. 
Using this fact we can write \eqref{eq:1Gcancel2} as:

\begin{align} \label{eq:1Gcancel3}
 \int &\df^{d-2}\mathbf{k}_T \, \df^{d-2}\mathbf{k'}_T  \, \delta(E_T = |\mathbf{k'}_T|+|\mathbf{k'}_T - \mathbf{q}_T |) \times \\ \nonumber 
& f_P(\mathbf{k}_T) f^*_P(\mathbf{k'}_T) \left[ L(\mathbf{k}_T \to \mathbf{k'}_T) +  L^*(\mathbf{k'}_T \to \mathbf{k}_T)\right]
\end{align}

Now $L+L^*$ can cancel as before, giving zero for the contribution from the Glauber region.

\begin{figure} 
\centering
\includegraphics[scale=0.6]{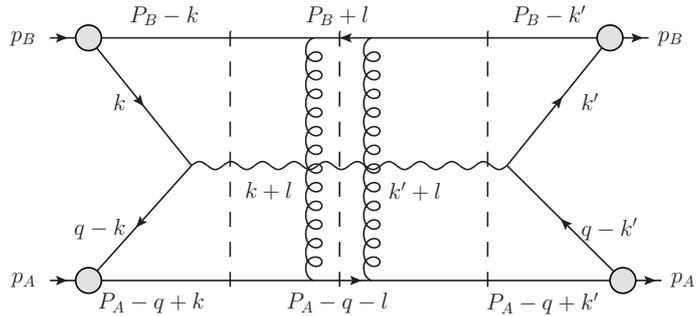}
\caption{Graph with two Glauber exchanges between the spectators.The three cuts that
allow the gluon to remain in the Glauber region are denoted on the graph as dashed
lines.}
\label{fig:twoglauber}
\end{figure}

So, we have seen that the Glauber region cancels for one Glauber gluon exchange both
for the MPI insensitive variable $p_T$, and for the MPI sensitive variable $E_T$, though 
considerably more gymnastics was required to show this for the latter case. Let us
now add in one more Glauber gluon, giving rise to the graph shown in figure \ref{fig:twoglauber}. 
It is once again simple to show that both gluons in this graph can be trapped in the
Glauber region (by taking a loop momentum looping round the left hand side for the
left-hand gluon, and a loop momentum looping around to the right for the right-hand
gluon), and that this region contributes at leading power of $\Lambda_{QCD}/Q$. Does
the Glauber region cancel for this graph as well?

\begin{figure} 
\centering
\includegraphics[scale=0.6]{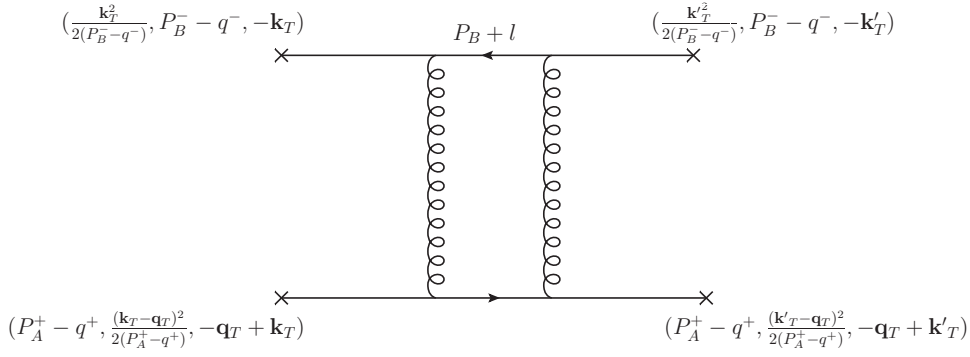}
\caption{`Glauber gluon' subgraph of the two Glauber exchange graph in figure 
\ref{fig:twoglauber}.}
\label{fig:twoglaubercut}
\end{figure}

There are now 3 cuts of the graph (`Glauber cuts') that allow both gluons to
remain in the Glauber region, which are drawn in figure \ref{fig:twoglauber}. We can follow the same 
reasoning as we did for the one-Glauber-exchange graph and factor a parton-model graph
from each cut graph, leaving us with the sum over (internal + external) physical cuts of the
2-gluon exchange graph depicted in figure \ref{fig:twoglaubercut}. By external cuts we mean
cuts running through external legs of the Glauber subgraph (we shall later refer to these also
as `absorptive' cuts), and by internal cuts we mean cuts running through internal legs of the Glauber
subgraph (these will later also be referred to as `real' cuts). If we just measure the total cross 
section (no measurement function) then the sum over cuts cancels for given 
$\mathbf{k}_T, \mathbf{k'}_T$. As mentioned before, the sum over the two external cuts
gives:
\begin{equation} \label{eq:2gexternal}
L(\mathbf{k}_T \to \mathbf{k'}_T;l) + L^*(\mathbf{k'}_T \to \mathbf{k}_T;l) = i\mathcal{M}(\mathbf{k}_T \to \mathbf{k'}_T;l) - i \mathcal{M}^*(\mathbf{k'}_T \to \mathbf{k}_T;l)
\end{equation}
where $L(\mathbf{k}_T \to \mathbf{k'}_T;l)$ [$\mathcal{M}(\mathbf{k}_T \to \mathbf{k'}_T;l)$] is the 
one-loop graph [amplitude piece] corresponding to figure \ref{fig:twoglaubercut}. The sole internal cut gives 
\begin{align} \label{eq:2ginternal}
&\int \df\Phi_2   L(\mathbf{k}_T \to \mathbf{l}_T) L^*(\mathbf{k'}_T \to \mathbf{l}_T) \\ \nonumber
= &\int \df\Phi_2   \mathcal{M}(\mathbf{k}_T \to \mathbf{l}_T) \mathcal{M}^*(\mathbf{k'}_T \to \mathbf{l}_T)
\end{align}
where $d\Phi_2$ is the two-particle on-shell phase space, as in equation \eqref{eq:phspgen},
and the $L$ and $\mathcal{M}$ here are the tree-level quantities.
However \eqref{eq:2gexternal} is equal to minus \eqref{eq:2ginternal} according to the diagram-by-diagram
version of \eqref{eq:GenOp} (i.e. the Cutkosky rules), so the sum of three cuts gives zero.
Also, if we have the MPI insensitive observable $p_T$ then all cuts contribute 
to the same value of the observable by momentum conservation (for arbitrary loop
transverse momentum $\mathbf{l}_T$), and cancel against each other just as occurs for the 
total cross section.

Now, what happens if we have the MPI sensitive observable $E_T$? Here the cancellation
fails because the `real' cut of the diagram between the Glauber gluons gives a value
of $E_T$ that depends on the transverse component of the `internal' loop momentum 
$\mathbf{l}_T$ (where by internal we mean internal to the $q\bar{q} \to q\bar{q}$
scattering process), whilst the other two `absorptive' cuts either side of the Glauber
gluons give an $E_T$ that does not depend on this internal loop momentum. 

One might wonder if there is some clever simultaneous change of variables involving both 
the `external' variables $\mathbf{k}_T$ and $\mathbf{k'}_T$ and the `internal' loop variable 
$\mathbf{l}_T$ that can be used to bring about the cancellation, as we found in the 
one-Glauber case. In this case we argue that no such variable transform is possible. Making a change in the
internal variable $\mathbf{l}_T$ is inadvisable, as the Cutkosky cancellation to which
we would like to appeal is a cancellation that occurs point by point in spatial momentum
(as is shown in Appendix B of \cite{Sterman:1995fz}). Therefore we should match the 
parameterisation of spatial momentum between the external and internal cuts of the graph.
This essentially leaves us with the relabelling $\mathbf{k}_T \leftrightarrow \mathbf{k'}_T$,
which in this case does not help (due to the internal cut).

Let us see this more explicitly. We consider the three cuts of figure \ref{fig:twoglaubercut} 
with this figure embedded in the larger diagram figure \ref{fig:twoglauber}:
\begin{align}
\int \df^{d-2}\mathbf{k}_T \, \df^{d-2}\mathbf{k'}_T  f_P(\mathbf{k}_T) f^*_P(\mathbf{k'}_T) \,  & \bigg\{\int \dfrac{\df^dl}{(2\pi)^d} \,  \big[ L(\mathbf{k}_T \to \mathbf{k'}_T;l) \delta(E_T = |\mathbf{k'}_T|+|\mathbf{k'}_T - \mathbf{q}_T |) \nonumber \\ 
        & + L^*(\mathbf{k'}_T \to \mathbf{k}_T;l)  \delta(E_T = |\mathbf{k}_T|+|\mathbf{k}_T - \mathbf{q}_T |) \big] \\ \nonumber
        + &\int \df\Phi_2   L(\mathbf{k}_T \to \mathbf{l}_T) L^*(\mathbf{k'}_T \to \mathbf{l}_T)  \delta(E_T = |\mathbf{l}_T|+|\mathbf{l}_T + \mathbf{q}_T |) \bigg\}
\end{align}

In the Glauber
region at leading power the numerator factor in the $L$ factors just gives a constant and is irrelevant to the 
present discussion. The graph $L(\mathbf{k}_T, \mathbf{k'}_T;l)$ is invariant under 
$\mathbf{k}_T \leftrightarrow \mathbf{k'}_T$ at leading power. This is because when we flip
$\mathbf{k}_T \leftrightarrow \mathbf{k'}_T$ we get back the original graph but with the directions
of all momentum arrows reversed. Flipping the directions of the momenta affects only the numerator,
but in the region we are interested in, the numerator is just a constant and is unchanged by the flip.
Using this, plus the previously established fact that $f_P^* = -f_P$, in the second term:
\begin{align}
\int \df^{d-2}\mathbf{k}_T \, \df^{d-2}\mathbf{k'}_T  \, f_P(\mathbf{k}_T) f^*_P(\mathbf{k'}_T) & \bigg\{ \int \dfrac{\df^dl}{(2\pi)^d} \,  \big[ L(\mathbf{k}_T \to \mathbf{k'}_T;l) \delta(E_T = |\mathbf{k'}_T|+|\mathbf{k'}_T - \mathbf{q}_T |) \nonumber \\ 
        & + L^*(\mathbf{k'}_T \to \mathbf{k}_T;l)  \delta(E_T = |\mathbf{k'}_T|+|\mathbf{k}'_T - \mathbf{q}_T |) \big] \\ \nonumber
        + &\int \df\Phi_2   L(\mathbf{k}_T \to \mathbf{l}_T) L^*(\mathbf{k'}_T \to \mathbf{l}_T)  \delta(E_T = |\mathbf{l}_T|+|\mathbf{l}_T + \mathbf{q}_T |) \bigg\}
\end{align}

Now $L$ and $L^*$ have the same value of the measurement and we can combine them together using the
Cutkosky rule:
\begin{equation}
\int  \dfrac{\df^dl}{(2\pi)^d} \,  \big[ L(\mathbf{k}_T \to \mathbf{k'}_T;l) + L^*(\mathbf{k'}_T \to \mathbf{k}_T;l) \big] = - \int \df\Phi_2   L(\mathbf{k}_T \to \mathbf{l}_T) L^*(\mathbf{k'}_T \to \mathbf{l}_T) 
\end{equation}
Then we obtain:
\begin{align} \label{twoglexp}
\int \df^{d-2}\mathbf{k}_T \, \df^{d-2}\mathbf{k'}_T   \, &f_P(\mathbf{k}_T) f^*_P(\mathbf{k'}_T) \int \df\Phi_2   L(\mathbf{k}_T \to \mathbf{l}_T) L^*(\mathbf{k'}_T \to \mathbf{l}_T)  \times  \\ \nonumber
         &  \big[ - \delta(E_T = |\mathbf{k'}_T|+|\mathbf{k'}_T - \mathbf{q}_T |) +   \delta(E_T = |\mathbf{l}_T|+|\mathbf{l}_T + \mathbf{q}_T |) \big]
\end{align}

The explicit expression for $\int d\Phi_2   L(\mathbf{k}_T \to \mathbf{l}_T) L^*(\mathbf{k'}_T \to \mathbf{l}_T) $ at leading power is as follows:
\begin{align} \label{eq:Lcut}
\int \df\Phi_2   L(\mathbf{k}_T & \to \mathbf{l}_T) L^*(\mathbf{k'}_T \to \mathbf{l}_T) \\ \nonumber
    &= \int \dfrac{\df^d l}{(2\pi)^d} \dfrac{\text{const.}}{(l+k)^2(l+k')^2}(2\pi)\delta((P_B+l)^2)(2\pi)\delta((P_A-q-l)^2) \\ \nonumber
    &\simeq \int \dfrac{\df^{d-2}\mathbf{l}_T}{(2\pi)^{(d-2)}} \dfrac{\text{const.}}{2(P_A^+-q^+)2(P_B^--q^-)(\mathbf{l}_T+\mathbf{k}_T)^2(\mathbf{l}_T+\mathbf{k'}_T)^2}
\end{align}
Due to the mismatch in the delta function arguments, \eqref{twoglexp} does not cancel to zero for
general $f_P(\mathbf{k}_T)$.

The lack of a cancellation for fixed $\mathbf{k}_T$ and $\mathbf{k'}_T$, which worked in the total
cross section and $p_T$ cases, can be illustrated by looking at a simpler observable which will shall
refer to as $T_n$. To obtain $T_n$ one
just measures the sum of squared transverse momenta of all non-$V$ particles in the $n$-hemisphere. 
For simplicity let us restrict ourselves for the moment to the 
case $\mathbf{k}_T = \mathbf{k'}_T$. Then the function inside the square brackets in \eqref{twoglexp}
is:
\begin{align} \label{eq:Tndelta}
[-\delta(T_n = \mathbf{k}_T^2) + \delta(T_n = \mathbf{l}_T^2)]
\end{align}

If $\mathbf{k}_T^2 \neq T_n$ then the first delta function in \eqref{eq:Tndelta} does not 
contribute. The second gives a finite result when inserted into the integral over $l$, equation
\eqref{eq:Lcut}:
\begin{align} \label{eq:Tnrealint}
\begin{cases} 
 \text{const.} \times \dfrac{T_n+\mathbf{k}_T^2}{(\mathbf{k}_T^2-T_n)^3} & \text{for } \mathbf{k}_T^2 > T_n \\
 \text{const.} \times \dfrac{T_n+\mathbf{k}_T^2}{(T_n- \mathbf{k}_T^2)^3} & \text{for } \mathbf{k}_T^2 < T_n \\
\end{cases}
\end{align}
with the constant equal in both cases. Note that this function appears to have an (infra-red) divergence as 
$\mathbf{k}_T^2 \to T_n$ -- however this divergence is cancelled by the integral associated 
with the absorptive cuts (first delta function in \eqref{eq:Tndelta}).

It is clear from \eqref{eq:Tnrealint} that the integral over $l$ for the observable $T_n$ does
not give zero for fixed $\mathbf{k}_T = \mathbf{k}'_T$. Even when integrating along
$\mathbf{k}_T = \mathbf{k}'_T$ we will not get zero unless the weight function 
$f_P(\mathbf{k}_T)f_P^*(\mathbf{k}'_T)$ happens to have exactly the right shape such
that the contribution from the absorptive cuts at $\mathbf{k}_T^2 = T_n$ can cancel the 
contribution from the real cuts distributed over all $\mathbf{k}_T^2$ (e.g. if 
$f_P(\mathbf{k}_T)f_P^*(\mathbf{k}'_T)=1$, which of course is not at all realistic).
Such a cancellation seems therefore implausible.

In the more general case $\mathbf{k}_T \neq \mathbf{k'}_T$ the maths is more complicated,
but the qualitative picture remains similar. The contribution from the central real
cut is the only nonzero one for $T_n \neq \mathbf{k}_T^2, \mathbf{k'}_T^2$. As either $\mathbf{k}_T^2$ 
or $\mathbf{k'}_T^2$ tend to $T_n$, this contribution becomes divergent, but the divergence
is cancelled by the contribution from the absorptive cuts, if we symmetrise the delta function
in this contribution: $\delta(T_n = \mathbf{k}_T^2) \to \tfrac{1}{2}[\delta(T_n = \mathbf{k}_T^2)
+ \delta(T_n = \mathbf{k'}_T^2) ]$ (we are allowed to do this due to the symmetry of the rest of 
the integrand). Let us write out the integral over $l$ explicitly, dropping the constant factors
and implementing the symmetrisation in the delta function:
\begin{align}
 \int \dfrac{\df^{d-2}\mathbf{l}_T}{(2\pi)^{(d-2)}} & \dfrac{1}{(\mathbf{l}_T+\mathbf{k}_T)^2(\mathbf{l}_T+\mathbf{k'}_T)^2} \\ \nonumber
                                                  & \times  [-\tfrac{1}{2}\delta(T_n = \mathbf{k}_T^2) - \tfrac{1}{2} \delta(T_n = \mathbf{k'}_T^2) + \delta(T_n = \mathbf{l}_T^2)]
\end{align}

Now we consider integrating $T_n$ over a small region around $T_n = \mathbf{k}_T^2$ (for
example). Only one of the two delta functions associated with the absorptive cuts gives
a nonzero contribution, which according to a straightforward calculation has the following
divergent piece:
\begin{equation} \label{eq:TnAbsDiv}
\dfrac{\pi}{(2\pi)^2\epsilon}\dfrac{1}{(\mathbf{k}_T-\mathbf{k}'_T)^2} + \mathcal{O}(\epsilon^0)
\end{equation}
where we take $d=4-2\epsilon$ to regulate the integral.

In the integral associated with the real cut, the divergence comes from the region around
$\mathbf{l}_T = -\mathbf{k}_T$. We can therefore replace the denominator factor 
$(\mathbf{l}_T+\mathbf{k'}_T)^2$ by $(\mathbf{k}_T-\mathbf{k'}_T)^2$ for the divergent
part. Then we get from the contribution from the real cut:
\begin{align} \label{eq:TnRealDiv}
&\dfrac{1}{(\mathbf{k}_T-\mathbf{k}'_T)^2}\int \dfrac{\df^{d-2}\mathbf{l}_T}{(2\pi)^{(d-2)}} \dfrac{1}{(\mathbf{l}_T+\mathbf{k}_T)^2} + \mathcal{O}(\epsilon^0) \\ \nonumber
= & -\dfrac{\pi}{(2\pi)^2\epsilon}\dfrac{1}{(\mathbf{k}_T-\mathbf{k}'_T)^2} + \mathcal{O}(\epsilon^0)
\end{align}
where the integral on the first line is only over a small region around $\mathbf{l}_T = -\mathbf{k}_T$.
We see that the divergent parts \eqref{eq:TnAbsDiv} and \eqref{eq:TnRealDiv} cancel.

We end up with some more complicated version of \eqref{eq:Tnrealint} with all 
divergences regularised. When we integrate over $\mathbf{k}_T$ and $\mathbf{k}'_T$ a very 
particular form for the parton model weight function $f_P(\mathbf{k}_T)f_P^*(\mathbf{k}'_T)$ 
would be required to achieve a cancellation -- such a cancellation is again not very plausible.

\begin{figure} 
\centering
\includegraphics[scale=0.6]{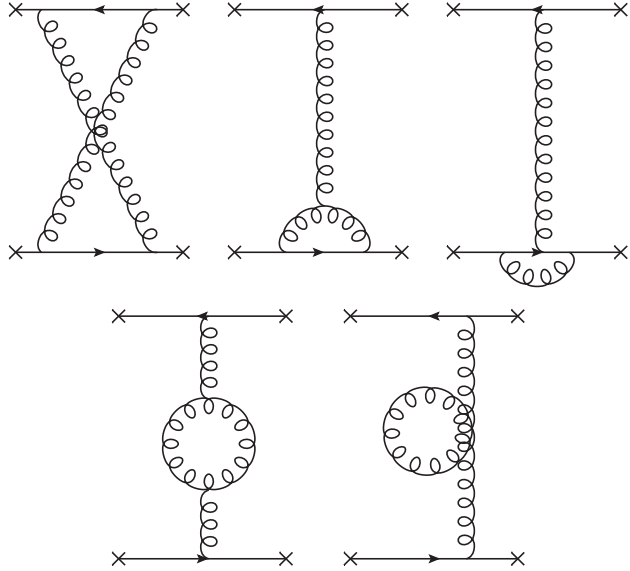}
\caption{Alternative two-gluon-exchange diagrams. We have not drawn diagrams that
are related to the ones above by reflection with respect to a horizontal axis.}
\label{fig:crossedgluon}
\end{figure}

One might wonder if it is still possible to cancel figure \ref{fig:twoglauber} 
against other graphs in which the `separate' $q\bar{q} \to q\bar{q}$ two-gluon 
exchange process is one of the processes depicted in figure \ref{fig:crossedgluon}.
We can see that this will not happen, because these graphs only have external or
absorptive cuts, and no real cuts. We can follow through the same logic as we did in
equations \eqref{eq:1Gcancel} to \eqref{eq:1Gcancel3}, and cancel the contribution of
these graphs to the $E_T$ spectrum individually. In these cases we obtain
$L(\mathbf{k}_T \to \mathbf{k}'_T;l) = L(\mathbf{k}'_T \to \mathbf{k}_T; l)$ using the same 
argument as was applied to figure \ref{fig:twoglaubercut}.

Another possibility one might think of is that figure \ref{fig:twoglauber} could be 
cancelled against other graphs containing central soft or collinear connections from 
the Glauber subgraph to the main scattering process. We have not analysed this 
possibility in detail but find it implausible for such a cancellation to occur. 
Very shortly we will see that the Glauber gluon process in figure \ref{fig:twoglauber} is 
intimately related to soft MPI in the hadron-hadron scattering process. The cancellation
of figure \ref{fig:twoglauber} by graphs with soft or collinear modes would in some sense
be equivalent to the suppression of soft MPI by soft or collinear interactions, which
we do not find very plausible. Note that in the inclusive cross section or $p_T$ of $V$
cases, the cancellation of figure \ref{fig:twoglauber} does not imply a suppression of soft MPI,
but an insensitivity of the observables to whether additional soft interactions occurred 
or not. We know from the above discussion that $E_T$ is certainly sensitive
to whether an additional scattering occurred or not.

So how does all of this discussion of factorisation breaking relate to MPI, and 
why do we refer to $E_T$ as MPI sensitive? This should be clear already from
our model calculation involving two Glauber gluons. Here, the process effectively splits 
up into two scattering processes -- the high-scale hard process, and a lower scale scattering 
process (for the cuts internal to the Glauber system) or absorptive interaction (for the 
external cuts to the Glauber system). The possibility of the real cut in the Glauber subgraph, 
or MPI, and the fact that the MPI changes the observable, is what results in the factorisation 
breaking.  Note that in the Monte Carlos one does not calculate the virtual graphs explicitly, 
but that they are included implicitly via unitarity constraints.

Note that with an interpretation of figure \ref{fig:twoglauber} as a high scale 
primary interaction producing $V$ plus a much lower scale secondary QCD interaction,
one might expect the lower scale interaction to happen at a smaller `shower time' in
the initial state shower than the hard interaction (where `shower time' represents the
virtuality of the partons in the shower and increases towards the hard interaction). 
This is the situation in (for example) the {\tt Pythia} initial state shower + MPI model
(see Figure 1 of \cite{Sjostrand:2004ef}). In \cite{Collins:1350496} an analysis of
the spacetime locations of the interactions is used to show that the additional interaction
actually occurs first in space-time as well as shower time.

\begin{figure} 
\centering
\includegraphics[scale=0.6]{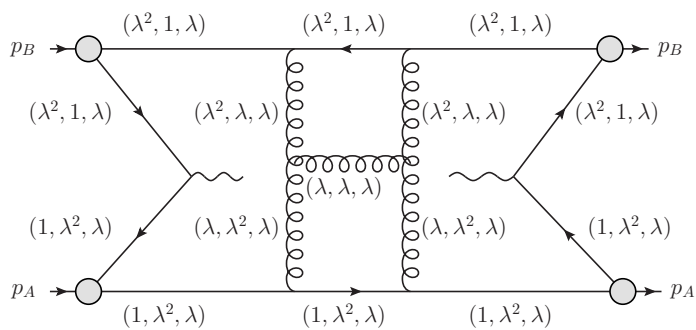}
\caption{Interaction between spectators composed from `collinear Glauber'
uprights and a central soft rung. The scaling of the momenta at the collinear
Glauber + central soft pinch is denoted on the figure. Each momentum $r_i$ on the 
uprights is now referred to as `collinear Glauber' because one of $|r_i^+|$ or
$|r_i^-|$ is now non-negligible compared to the transverse momentum, but the 
momentum satisfies the basic condition for a Glauber momentum $|r_i^+ r_i^-| \ll \mathbf{r}_{Ti}^2$.}
\label{fig:twoglauberplussoft}
\end{figure}

Many models of MPI in Monte Carlo programs -- notably the ones in {\tt Pythia}
\cite{Sjostrand:1987su, Sjostrand:2004pf, Sjostrand:2004ef}, {\tt Herwig}(++) \cite{Borozan:2002fk, 
Bahr:2008dy,Bahr:2008pv}, and the {\tt Pythia}-inspired
{\tt AMISIC++} model in {\tt Sherpa} \cite{Alekhin:2005dx} -- take additional 
interactions to be just extra $2 \to 2$ scattering processes. It may be important
to consider more general $2 \to n$ processes. Let us go back for the moment
to our model, and consider the next step up in complexity from our `two 
Glauber gluon' example. In this example we make the simplest gluon ladder possible
by inserting a horizontal gluon `rung' between the vertical gluons in
figure \ref{fig:twoglauber}, to obtain figure \ref{fig:twoglauberplussoft}. This graph can
have a leading pinch in the region in which the vertical gluons are in the `collinear 
Glauber' region, and the rung is central soft -- the scalings of all the particles 
at the pinch are demonstrated on figure \ref{fig:twoglauberplussoft}. In the $E_T$ 
case, we again do not achieve a cancellation between the central `real' cut and the
`absorptive' cuts to either side of the Glauber subgraph, as we also found for the 
simpler two-Glauber-gluon graph. 

Figure \ref{fig:twoglauberplussoft} is suppressed by an additional power of $\alpha_S$, but it is 
known from BFKL physics that when the rapidities are strongly ordered (as
indeed they are in figure \ref{fig:twoglauberplussoft}) then one picks up an additional enhancement
from rapidity logarithms that acts to counterbalance the $\alpha_S$
suppression (see for example \cite{KovchegovQCD}). Indeed, we can actually insert arbitrary numbers of gluon 
rungs into the Glauber process, forming something akin to a Pomeron, and
still be at leading logarithmic order in the BFKL sense (the connection between
Glauber gluons and the Pomeron/BFKL physics is also highlighted in \cite{Collins:1350496, Forshaw:2012bi, StewartTalk2, RothsteinTalk, Fleming:2014rea}). 
These objects, as well as
other effects from Pomeron physics such as Pomeron splitting
and merging, are not included in e.g. default {\tt Pythia} and {\tt Herwig} (except
in the parts of these programs designed to describe diffractive observables), but 
could well be important. Due to this fact we cannot be completely confident 
that the contribution to $E_T$ from the underlying event generated by {\tt Herwig++}
in the studies \cite{Papaefstathiou:2010bw, Grazzini:2014uha} will be completely accurate. We note here that the
Monte Carlo codes {\tt SHRiMPS} \cite{Martin:2012nm} and {\tt DIPSY} \cite{Flensburg:2011kk} do
incorporate some of the necessary BFKL/Pomeron-type effects.

\subsection{Other Observables} \label{sec:OtherObs}

With the connection between standard factorisation violation, two Glauber exchange and soft MPI, we 
expect any observable sensitive to MPI $O_S$ not to obey factorisation when the value of that observable
is parametrically small compared to the hard scale (where here again we have taken $O_S$ to have mass
dimension $1$). Consider for example beam thrust. In this case the function inside the square brackets 
in \eqref{twoglexp} becomes:
\begin{align} \label{eq:beamthrustdeltas}
\bigg[ &-\delta\left(b_a^+ =  \dfrac{(\mathbf{k}'_T-\mathbf{q}_T)^2}{2(P_A^+-q^+)}\right)\delta\left(b_b^+ = \dfrac{\mathbf{k'}_T^2}{2(P_B^--q^-)}\right)  \\ \nonumber
&+ \delta\left(b_a^+ = \dfrac{(\mathbf{l}_T+\mathbf{q}_T)^2}{2(P_A^+-q^+)}\right)\delta\left(b_b^+ = \dfrac{\mathbf{l}_T^2}{2(P_B^--q^-)}\right)  \bigg]
\end{align}
with $B_a^+ = \sqrt{2}b_a^+$, $B_b^+ = \sqrt{2}b_b^+$. We can see that also in this case the integrals
associated with the real and absorptive cuts will not cancel.\footnote{Imagine if we add some initial state
radiation (ISR) to the `primary' $q\bar{q} \to V$ process in figure \ref{fig:twoglauber}. The effect of this initial
state radiation is what is resummed in the standard factorisation formula. Then if $x_p$ corresponds to the 
$x$ values in the primary interaction, then one expects the contribution to the beam thrust from this ISR to
be roughly of order transverse momentum squared divided by $x_pP$ (with $P \sim P_A^+ \sim P_B^-$). By contrast
the contribution from the Glauber miscancellation in \eqref{eq:beamthrustdeltas} is of order transverse momenta squared
divided by $x_sP$, with $x_s$ the $x$ values taken by the spectators (and we have $x_s+x_p = 1$). If  
$q^+,q^- \ll P_A^+,P_B^-$ (as is often the case for processes measured at the LHC) such that $x_p \ll 1$, and
$x_s \gg x_p$ in our model, then one might be tempted to argue that the Glauber miscancellation for beam thrust
is energy suppressed. However, this is neglecting the fact that in reality in the proton we can have many spectators, 
and also have have ISR from the spectators, such that the $x$ values of the spectators can be (and often are)
as small or smaller than the $x$ values for the primary interaction. So one cannot argue that the Glauber 
miscancellation is energy suppressed for beam thrust in this way.}

It is particularly interesting in the case of beam thrust to consider figure \ref{fig:twoglauberplussoft}.
For beam thrust the central real cut of figure \ref{fig:twoglauberplussoft} gives a power suppressed 
contribution compared to the absorptive cuts (which are leading power), so in this case the lack of a cancellation 
between real and absorptive cuts becomes particularly clear. The reason why the real cut gives a power suppression 
for beam thrust is that its contribution to beam thrust is spread over a much larger phase space than that from the 
parton model or absorptive graphs, being of order $\lambda Q$ rather than $\lambda^2 Q$.

One example of a further observable which can be considered to be MPI sensitive is transverse thrust. This is defined 
by computing $\sum_i |\mathbf{q}_{iT} \cdot \mathbf{n}_{iT}| / \sum_i |\mathbf{q}_{iT}|$ and then maximising over the 
direction of the transverse vector $\mathbf{n}_{iT}$. One can verify by following the steps
above that the CSS argument does not go through for this variable either. In fact we have experimental
data for this observable in hadron-hadron collisions \cite{Aaltonen:2011et, Aad:2012fza, Khachatryan:2011dx, 
Chatrchyan:2013tna}. In the CDF study \cite{Aaltonen:2011et} it was shown that an NLO+NLL result \cite{Banfi:2010xy} 
from the standard factorisation formula for this observable disagreed strongly with the shape of the
transverse thrust from data, and that the Monte Carlo curves including MPI agreed much better with the
data. This might be regarded as experimental evidence that the standard factorisation does not work
for this observable.

Another class of observables that is interesting to consider is the set of jet based observables -- e.g. 
the $p_T$ or mass of an individual jet. If we denote the momentum of the jet after the clustering algorithm
has been applied as $p_{J}$, then the $p_T$ of the jet is $\mathbf{p}_{TJ}$ and the mass squared of the jet 
is $p_J^2$. Another example is $\mathcal{T}_j$, which is given by identifying jets $j(R)$ of size $R$, and then taking the beam 
thrust of the jet which gives the largest contribution to this quantity \cite{Tackmann:2012bt}. For these variables, the equivalent to \eqref{twoglexp} has 
a sum of extra delta functions in each term expressing the fact that at least one of the emitted particles in 
the MPI or absorptive process should be inside the jet. For some general jet observable the contributions 
associated with the real and absorptive cuts in figure \ref{fig:twoglauber} will again not cancel, so once
again we get a $\mathcal{O}[(\lambda/Q)^0]$ contribution from the Glauber region/MPI and we cannot claim that
the standard factorisation formulae for these observables holds at leading power in $\lambda/Q$ either.
However, for jet observables, we have another parameter, the jet radius $R$. In order for the Glauber miscancellation to happen,
at least one of the absorptive or real sets of delta functions in the equivalent to \eqref{twoglexp} must be 
satisfied, a possibility that gets rarer and rarer the smaller the jet radius $R$ is. So for jet-based observables 
the Glauber miscancellation is suppressed by the jet size $R$ (as opposed to $\lambda/Q$), and these observables 
are less MPI sensitive such that we may usefully apply the standard factorisation formulae to predict them (provided
$R$ is not too large). The physical picture corresponding to this is clear -- for jet-based observables, we only 
collect up MPI over the area of the jet(s), and so are much less sensitive to them than we would be for the global 
observables discussed above (with the sensitivity decreasing as the jet radius $R$ decreases). The notion of MPI 
sensitivity in jet-based observables has been discussed before, in \cite{Dasgupta:2007wa}.

\subsection{MPI sensitive variable of same order as the hard scale} \label{sec:ObsSimHard}

In our discussion so far we did not specify the size of the MPI sensitive observable $O_S$. When the MPI sensitive 
variable is parametrically small with respect to the hard scale $Q$ (e.g. of order $\lambda Q$ in the case of $E_T$ 
or of order $\lambda^2 Q$ in the case of $B^+_{a,b}$) then the effect of the Glauber gluon miscancellation is
large. On the other hand, naive application of the CSS (or more recent Aybat-Sterman \cite{Aybat:2008ct}) argument
might lead one to believe that when one considers $O_S$ to be of values of order $Q$, then the standard factorisation
formula should apply (this is in particular for the cumulant distribution in $O_S$, where by the term `cumulant' here we mean 
the integral of the $O_S$ distribution from zero up to some value of $O_S$). Then one seems to be `inclusive'
up to the hard scale $Q$, and a standard factorisation formula should then be satisfactory. When $O_S \sim Q$, then 
the Glauber miscancellation in figure \ref{fig:twoglauber} only smears the observable by some power suppressed amount, 
which especially in the cumulant is not a significant effect.

However, the problem in practice is that an MPI sensitive observable $O_S$ can receive a cumulative contribution from many 
additional soft scatters, and there is the possibility of Glauber miscancellations on multiple spectator legs 
adding together to disrupt (broaden) the $O_S$ distribution significantly for not too large $O_S$ (even when 
it is of order $Q$). In this sense the $O_S$ distribution/cumulant can be sensitive to whether much lower scale
scatterings occur or not (i.e. not be sufficiently inclusive) even for $O_S\sim Q$.

This possibility of multiple additional interactions is built into the Monte Carlo generators. The effect of
MPI on the MPI sensitive observable $E_T$ has been studied using the Monte Carlo generator {\tt Herwig++} in
\cite{Papaefstathiou:2010bw, Grazzini:2014uha}. From section 5.3 of \cite{Papaefstathiou:2010bw} and section
4.3 of \cite{Grazzini:2014uha} we can see that the $E_T$ differential distribution (and also infer that the cumulant) changes
drastically when we add MPI according to the {\tt Herwig++} model, even when $E_T$ is of order $Q$ (and beyond).
This is when $Q$ is of order of the typical electroweak scales that are probed at the LHC, and no minimum $p_T$
cut on detected hadrons is imposed. By imposing some minimum $p_T$ cut on detected hadrons of the order of 
$1-2$ GeV (and thereby effectively becoming insensitive to the lowest scale MPI), the shape of the $E_T$ distributions 
becomes much narrower such that the cumulant in $E_T$ agrees better with the result from the factorisation formula 
when integrating up to $Q$. However, one could argue that this is not terribly useful since the value they are then 
agreeing on is the total cross section for $p + p \to V + X$, which we know we can predict via a factorisation formula. 
Also the strong sensitivity of the $E_T$ distribution/cumulant on a very low scale $p_T$ cut off on hadrons
shows that the $E_T$ distribution/cumulant is strongly sensitive on very much lower scale MPI
occurring, even when $E_T \sim Q$. Or, to put it another way, the $E_T$ distribution/cumulant is not a sufficiently
inclusive observable even when $E_T \sim Q$.

One might expect that there is some limit to how much $E_T$ can be generated by the MPI, since one cannot have an infinite number of
extra interactions (that generate enough $p_T$ to be measurable) and each extra interaction tends to only contribute a small amount 
to the observable. Then if one measured $E_T$ at values much larger than this limit one would expect to get a good description 
from the standard factorisation formula (up to some smearing of the order of the maximum value of the observable you can obtain from 
MPI). However, in the study \cite{Papaefstathiou:2010bw} it was observed that the MPI in {\tt Herwig++} are capable of generating an 
$E_T$ of $200-300$ GeV (this is consistent with the statement above that $E_T$ spectra for $E_T \sim Q$ were being strongly disrupted in 
{\tt Herwig++} even with $Q = \mathcal{O}(100\text{GeV})$. Taking this as a guideline we can see that at least for the MPI sensitive
observable $E_T$, finding a regime where the standard factorisation formula can be applied at the LHC is very difficult.

Due to the preference of additional interactions for small values of the 
observable, we expect typical events that contribute to the observable to contain
a large number of MPI. This means it is not possible to simply apply the 
framework being developed for double parton scattering (DPS) \cite{Diehl:2011yj, Manohar:2012jr, Blok:2011bu, Ryskin:2011kk, 
Gaunt:2011xd, Gaunt:2012dd} to this problem. To put it another way, MPI sensitive observables are
in general not a good tool to measure DPS because of strong pollution of the observable by MPI with much 
higher numbers of scatters.

\section{Conclusions}

We have shown explicitly that the classic Collins-Soper-Sterman (CSS) method of cancelling the
contribution from non-factorising Glauber gluons (or a straightforward extension thereof) fails at the level of 
two Glauber gluons for hadronic transverse energy $E_T$, when $E_T$ is much less than the hard scale of the primary
interaction $Q$. By connecting diagrams with such Glauber gluon exchanges to events with soft MPI, we argued that this 
failure of the CSS technique corresponds to an actual breakdown in the factorisation of the cross section for these 
observables into hard, collinear, and central soft functions at leading power. Using the same connection, we also 
argued that such factorisation fails for a larger class of variables, when these observables are $\ll Q$. We refer 
to these observables as MPI sensitive observables -- they are the ones whose distributions in hadron-hadron collisions 
can be disrupted strongly by the cumulative contribution of a large number of soft quasi-uncorrelated multiple parton 
interactions. Other examples of MPI sensitive variables are beam thrust $B^+_{a,b}$ and transverse thrust.

We suggested that since MPI sensitive observables can receive a cumulative contribution from MPI, there might be a 
problem in applying the standard `hard$\otimes$collinear$\otimes$central soft' factorisation to these observables
even when they are of order of the hard scale. The Monte Carlo studies of (for example) the $E_T$ spectrum when
$Q = \mathcal{O}(100\text{GeV})$ \cite{Papaefstathiou:2010bw, Grazzini:2014uha} give results consistent with this expectation.

Describing MPI sensitive variables correctly is a difficult problem. On the one hand, for a given value of the 
observable, there should be a significant contribution from events with a large number of very soft MPI.
Monte Carlo MPI models include these effects at some level (although of 
course the description of the very soft scatters can only be a model). On the other hand, MPI sensitive 
variables can be affected not just by $2 \to 2$ MPI but also $2 \to n$ scatters and a whole range of 
related effects familiar to the BFKL (or high energy) community. These effects are not described so
well by default {\tt Pythia} and {\tt Herwig} (which just have just have $2 \to 2$ MPI) but are included to some
degree in other Monte Carlo codes such as {\tt SHRiMPS} and {\tt DIPSY}. Including the full range of these effects
is of course highly nontrivial.

An alternative approach is of course to avoid MPI sensitive observables. Global observables such as the ones
we have discussed tend by their nature to be MPI sensitive (because they collect up all the particles produced in the
event, including the huge number of soft MPI distributed more or less evenly over the event). Jet-based 
observables are much less `MPI sensitive' in this regard, because they only collect particles from MPI over the
area of the jets, which is much smaller than the whole area of the event in general. This is of course the
strategy that has been favoured at the LHC and other hadron colliders. One should be cautious not 
to use jet areas that are too large -- then the observable starts to become more MPI sensitive \cite{Dasgupta:2007wa}.

\begin{acknowledgments}
The author of this paper gratefully acknowledges conversations with Markus Diehl, Zoltan Nagy, Stefan Prestel, Maximilian Stahlhofen, and Frank Tackmann, 
and thanks Markus Diehl and Frank Tackmann for many useful comments on the manuscript.
The Feynman diagrams in this paper have been drawn using {\tt JaxoDraw}~\cite{Binosi:2008ig}.
\end{acknowledgments}

\appendix
 
\section{Generalised Optical Theorem} \label{sec:GenOp}

Here we briefly review the most simple derivation of the generalised optical theorem (see 
for example \cite{Peskin:1995ev}). The Cutkosky
rules \cite{Cutkosky:1960sp, Veltman:1994Ve, Sterman:1995fz} are a more generalised form of this relation.

From the unitarity of the scattering matrix $S$ we can infer that $S$ must satisfy:
\begin{align} \label{eq:SSdag}
S^{\dagger}S = \mathbb{I}
\end{align}

We separate $S$ as usual into the identity matrix $\mathbb{I}$ (which corresponds to the particles 
passing through each other with no scattering) and a transition matrix $T$ (which always appears
accompanied by the factor $i$):
\begin{align} \label{eq:SandT}
S = \mathbb{I} + iT
\end{align}

Substituting \eqref{eq:SandT} into \eqref{eq:SSdag} we get:
\begin{align} \label{Cutkosky1}
iT -iT^\dagger + T^\dagger T = 0
\end{align}

Now we sandwich \eqref{Cutkosky1} between two different states $\bra{f}$ and $\ket{i}$. We use
the fact that:
\begin{align}
\bra{f}T\ket{i} = i(2\pi)^d \delta^{(d)}(p_i = p_f) \mathcal{M}(i \to f)
\end{align}
and:
\begin{equation}
\bra{f}T^\dagger\ket{i} = \bra{i}T\ket{f}^*
\end{equation}
to obtain:
\begin{align} \label{eq:GenOp}
i\left[\mathcal{M}(i \to f) - \mathcal{M}^*(f \to i)\right] = -\sum_X \int \df\Phi_X \mathcal{M}(i \to X) \mathcal{M}^*(f \to X) 
\end{align}
where $X$ is any possible intermediate state, and $d\Phi_X$ is the on-shell phase space element for
$X$. Denoting the momenta in $X$ as $p_j$, this is given by:
\begin{align} \label{eq:phspgen}
\df\Phi_X = (2\pi)^d\delta^{(d)}\left(p_{i(f)} - \sum_j p_j\right)\prod_j \dfrac{\df^dp_j}{(2\pi)^d} (2\pi) \delta\left(p_j^2-m_j^2\right) 
\end{align}

The relation \eqref{eq:GenOp} is the generalised optical theorem. The Cutkosky rules are more general than this, and state
that \eqref{eq:GenOp} applies also on a diagram-by-diagram basis (with the sum over $X$ being replaced by a sum over physical
cuts of the graph), and point-by-point in the spatial momenta of all (external+loop) particles \cite{Cutkosky:1960sp, Veltman:1994Ve, Sterman:1995fz}.

\bibliographystyle{jhep}
\bibliography{factbreak}

\end{document}